\documentclass[superscriptaddress,amsmath,amssymb,aps,epsf,prl,twocolumn]{revtex4-2}
\usepackage[T1]{fontenc}
\usepackage{bm}
\usepackage{amsmath}
\usepackage{amsfonts}
\usepackage{amssymb}
\usepackage[dvipsnames]{xcolor}
\usepackage{hyperref}
\usepackage{graphicx}

\usepackage{braket}

\begin{document}
	\title{Transient localization from the interaction with quantum bosons}
	\author{Hadi Rammal}
	\affiliation{Universit\'e Grenoble Alpes, CNRS, Grenoble INP, Institut N\'eel, 38000 Grenoble, France}
	\author{Arnaud Ralko}
	\affiliation{Universit\'e Grenoble Alpes, CNRS, Grenoble INP, Institut N\'eel, 38000 Grenoble, France}
	\author{Sergio Ciuchi}
	\affiliation{Dipartimento di Scienze Fisiche e Chimiche, Universit\'a dell’Aquila, Coppito-L’Aquila, Italy}
	\author{Simone Fratini}
	\affiliation{Universit\'e Grenoble Alpes, CNRS, Grenoble INP, Institut N\'eel, 38000 Grenoble, France}

\begin{abstract}

We carefully revisit the electron-boson scattering problem, going beyond popular semi-classical treatments. By providing numerically exact results valid at finite temperatures, we demonstrate the existence of a regime of electron-boson scattering where  quantum localization processes become relevant despite the absence of extrinsic disorder. Localization in the Anderson sense is caused by the emergent randomness resulting from a large thermal boson population,  being effective at transient times before diffusion can set in. Compelling evidence of this transient localization phenomenon is provided by the observation of a distinctive displaced Drude peak (DDP) in the optical absorption and the ensuing suppression of conductivity. Our findings identify a general route for anomalous metallic behavior that can broadly apply in interacting quantum matter.

\end{abstract}

\maketitle

\noindent

\paragraph{Introduction.---} 
Unusual charge transport properties are both a defining feature and an open challenge in quantum materials. Bad metals exhibit anomalously large resistivities, defying the very assumptions of  textbook, Bloch-Boltzmann, transport theory. Posing even more fundamental challenges, the  celebrated Fermi-liquid theory of metals  breaks down altogether in strange metals.  Several theoretical routes are being explored and novel concepts have been put forward to explain these anomalies.  In one class of scenarios,  the anomalies are ascribed to the unusual properties of the scatterers involved, that are able to overcome and replace the traditional electron-electron and electron-phonon scattering channels. 
Notable examples in this category include various types of critical, soft bosons appearing at quantum phase transitions or in extended critical phases \cite{Millis,Lohneysen,Collignon-STO}, eventually leading to bad metal behavior, strange metal behavior, or both. In a second group of scenarios, it is the very existence of electronic quasiparticles that is  questioned: the observed anomalies would then originate from the strange nature of the current carriers, that cannot be described as individual quasiparticles in the usual sense. 
Examples of these scenarios include  the reported destruction of quasiparticles by strong electron-electron interactions in correlated metals at high temperatures \cite{resilient,Vucicevic-vtx,Devereaux,spin-Holstein}, shifting the focus on the consequences of interaction-induced randomness   \cite{SYK-review}, or more radical model descriptions where the quasiparticle concept is abandoned from the outset \cite{Zaanen,Hartnoll-Lucas-Sachdev}.

In this Letter we explore a third route, where neither the scatterers nor the carriers are anomalous; what is anomalous instead is the nature of the scattering process itself, that will eventually lead to unusual charge transport.  What we propose here is that whenever there is emergent randomness in the problem,  it might become relevant to consider the effects of such randomness in full, including  the quantum localization phenomena that are traditionally associated with disordered systems. 

The idea that quantum localization processes can lead to anomalous charge transport at room temperature was initiated in recent years in order to explain the puzzling  properties of crystalline organic semiconductors \cite{Troisi06,PRBR11,AdvMat16}. There, just like in bad metals, the apparent time separating two subsequent  collision events is too short to be compatible with  semi-classical charge transport. The origin of the anomaly and the solution to the puzzle are  now understood as follows: in  molecular crystals the intermolecular vibrations are extremely soft for structural reasons, and they are therefore easily thermally populated; when these incoherent vibrations couple to the electronic motion, they result in strong randomness that is able to localize the electrons {\em à la} Anderson, albeit only on relatively short timescales. Electronic transport is then characterized by subsequent localization and delocalization processes  \cite{Thouless} that eventually lead to reduced diffusion, i.e. bad conduction.  This phenomenon comes with an associated fingerprint:  the electronic optical response exhibits a distinctive displaced Drude peak, providing direct evidence for localization.

Given the amount of existing knowledge on electron-phonon interactions, it came as a surprise that a previously unreported regime of electron-boson scattering could be uncovered. Even more puzzling, the latter has been shown to appear already for weak interactions, yet it had eluded even the most refined semi-classical  treatments available \cite{Fratini-PRL09}, revealing the need to fully account for the quantum nature of the electronic carriers and the  interference processes associated with it.

The phenomenon described above, now known as {\em transient localization}, was originally found  through direct 
time-dependent solution methods that treated the dynamic disorder classically \cite{Troisi06}. The  nature of  the observed localization was then understood analytically,  through a relaxation time approximation \cite{PRBR11,AdvMat16,SciPost}
building on the observation that slow dynamic scatterers  should behave increasingly close to static disorder as their characteristic frequency scale approaches zero.  More refined quantum-classical approaches have successfully been applied since, providing detailed, material-oriented predictions  \cite{Fratini2017,Giannini-NatComm19,Shuai-NComms21,Giannini2023,Shuai2023}.  Yet, whether and how this phenomenon can be sustained when the quantum nature of the bosons is restored has remained an open question: according to the common wisdom, quantum bosons should either bind to the charge carriers or scatter them inelastically, not localize them.

Hinting at the existence of transient localization  beyond the classical boson limit, signatures compatible with it have  been found in  fully-quantum numerical studies \cite{Fehske,Mishchenko,DeFilippis-PRL15,Shuai-NComms21,Jansen,Jankovic23}. It is the purpose of the present work to show systematically how  localization processes progressively develop from  electron-boson scattering through the emergence of a displaced Drude peak, and how these are able to suppress the carrier diffusion, therefore favoring bad conduction.

\paragraph{Model and method.---}
We consider the  Holstein model:
\begin{equation}
    H = -t \sum_{\langle i j \rangle} c_{i}^{\dagger}c_{j}  + \omega_0\sum _i a_i^{\dagger}a_i 
     + g \sum _i c_{i}^{\dagger}c_{i}(a_{i}^{\dagger}+a_{i}) \label{eq:Holstein}
\end{equation}
describing the local interaction of tight-binding electrons,  $c_i$,$c_{i}^{\dagger}$ at site $i$, with quantum bosons, $a_i$,$a_{i}^{\dagger}$, of characteristic frequency $\omega_0$.  In the original electron-phonon problem, the latter  represent local atomic or molecular vibrations; more generally, Eq. (\ref{eq:Holstein}) can effectively describe the interaction with collective or critical modes emerging from electron-electron interactions.  Because the  localization mechanism of interest here is active already at the single electron level, we focus on the 
scattering problem for an individual electron and set aside other explicit many-body effects. We define the dimensionless coupling strength $\lambda=g^2/(2t\omega_0)$ and set $\hbar=1$.

We solve the model Eq. (\ref{eq:Holstein}) using the finite temperature Lanczos method (FTLM) \cite{FTLM_1,prelovsek}  on finite-size chains of length $N_s$,  truncating the infinite Hilbert space of the bosons to a total maximum number $N_{bos}$ of quanta   on the chain (details in SM). 
This method provides a formally exact high-temperature expansion for both the thermodynamic quantities and dynamic correlation functions
down to temperatures of the order of a fraction of $t$.  Working at nonzero temperatures drastically reduces finite size effects as compared with $T=0$ exact diagonalization \cite{prelovsek}. These are further minimized by averaging over $N_\phi$ twisted boundary conditions (TBC) \cite{TBC_1,Gros1992}, increasing the effective number of allowed momenta from the nominal value $N_s$ to  $N_s \times N_\phi$. 
TBC sampling permits us to obtain reliable results down to the weak coupling regime, where the discrete nature of the non-interacting spectrum on small clusters is most critical. We use $N_s=4,5,6$ and $N_\phi=40$, sufficient for convergence in the temperature range of interest.

\begin{figure}[h]
        \includegraphics[width=8.4cm]{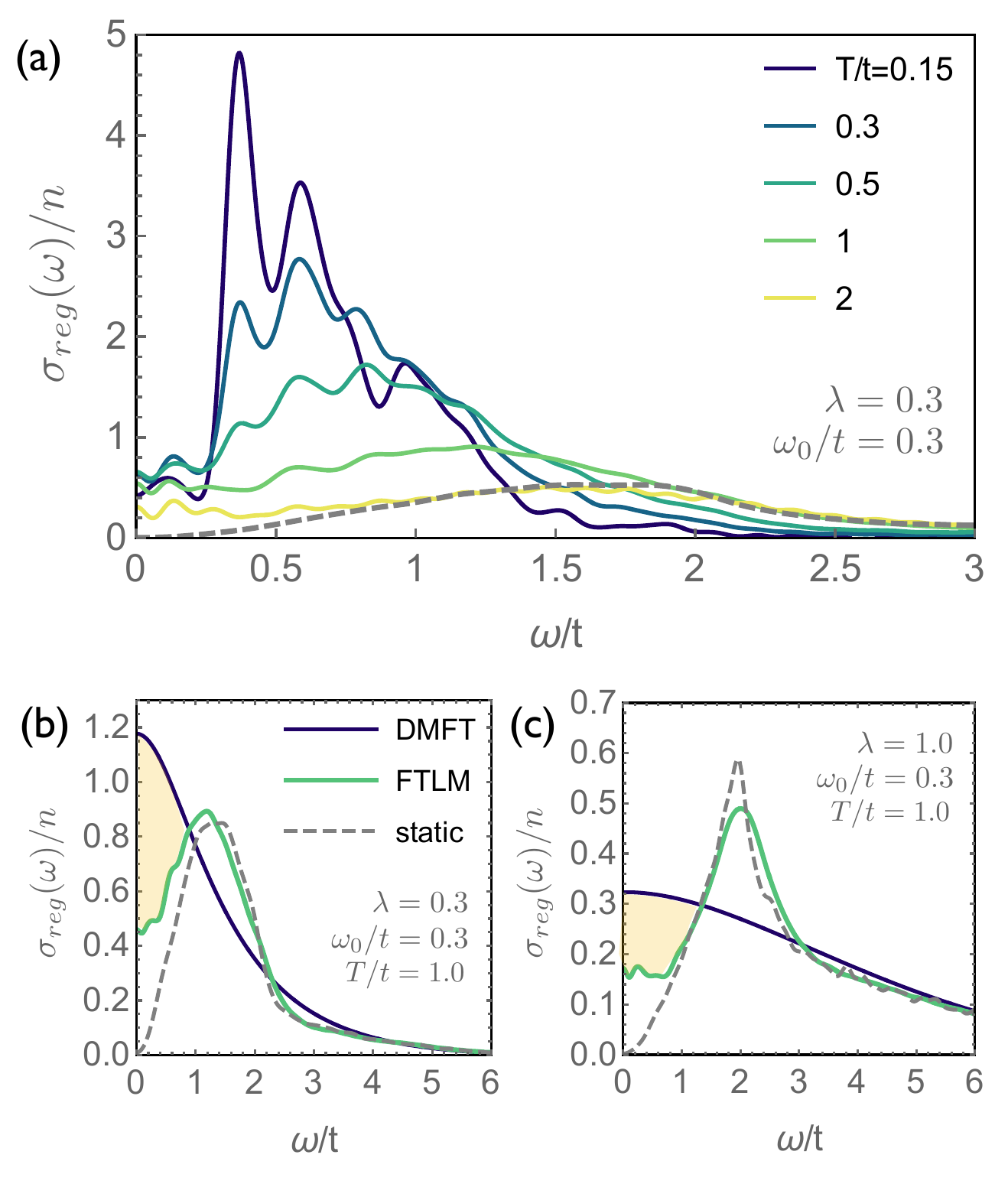}
\caption{(a) Regular part of the optical conductivity per particle for $\lambda=0.3$, $\omega_0/t=0.3$ calculated at different temperatures, expressed in units of $e^2a^2/\hbar$, with $a$ the lattice spacing (values $\sigma/n<2$ imply apparent scattering times shorter than the hopping time $t^{-1}$\cite{AdvMat16}). The cluster parameters are $N_s=4$, $N_{bos}=70$ and $N_{\phi}=40$ . The dashed line is the result from the static boson approximation at $T/t=2$ (a Gaussian filter of width $\delta=0.045t$ has been applied to all curves). The observed range of peak frequencies $\omega_L/t = 0.4 - 2$ corresponds to localization lengths $L=1 - 2.2$, consistently smaller than the system size. 
(b) Zoom of the FTLM data at $T/t=1.0$ compared with the DMFT result and the static boson approximation. (c) Similar to (b), at $\lambda=1.0$. 
In (b) and (c) the filter is $\delta=0.08t$.} 
    \label{fig:spectra-vsT}
\end{figure}

\paragraph{Emergent randomness and displaced Drude peak.---}
Fig. \ref{fig:spectra-vsT}(a) shows the regular ($\omega>0$) part of the optical conducitvity per particle calculated at different temperatures in the slow-boson (adiabatic) regime, $\omega_0/t=0.3$, for a moderate coupling strength $\lambda=0.3$ 
(see the SM for details). At low temperature, $T=0.15t<\omega_0$, the optical absorption recovers the expected weak-coupling picture \cite{optcondDMFT}: the spectrum is constituted of a main absorption peak at $\omega=\omega_0$ corresponding to single-boson emission, followed by weaker shakeoff replicas at multiples of $\omega_0$. Upon increasing the temperature, however, the finite-frequency absorption peak is not washed out as predicted by semi-classical approaches \cite{optcondDMFT}. Quite on the contrary, the peak at $\omega_0$  shifts to higher frequencies, progressively evolving into a broader shape (see also Ref. \cite{Mishchenko}). As we show next, this indicates that the nature of the electron-boson scattering smoothly evolves from independent, one-boson emission events, to a fundamentally different regime where 
self-generated disorder from abundant thermally excited bosons creates the conditions for localization in the Anderson sense.

To verify this hypothesis, we rewrite the harmonic Hamiltonian in first quantization as $\omega_0 (a^\dagger_ia_i +1/2)= P_i^2/2M + M \omega_0^2 X_i^2/2$ and explicitly neglect the dynamic part, formally taking the static boson limit $M\to \infty$,  $\omega_0\to 0$ with $M\omega_0^2= \ $constant \cite{SciPost}. The interaction part then becomes $\sum_i (g^\prime X_i) c_{i}^{\dagger}c_{i}$,  resulting in a one-body problem with disordered site energies $\epsilon_i=g^\prime X_i$ (the latter obey a Gaussian distribution of variance $\Delta=\sqrt{2 \lambda \omega_0/\tanh{(\omega_0/2T)}}$, which follows directly from the properties of the harmonic oscillator).  The corresponding optical conductivity is reported as a gray dashed line in Fig. \ref{fig:spectra-vsT}(a) for $T/t=2$: it shows  a disorder-induced localization peak at a frequency $\omega_L\simeq 2t/L^2$ \cite{PRBR11,AdvMat16,SciPost}, corresponding to the typical level spacing in a localization well of size $L$. 

A more stringent comparison with the static boson result is provided in Figs. \ref{fig:spectra-vsT}(b) and (c) for  moderate ($\lambda=0.3$) and strong ($\lambda=1.0$) electron-boson interactions.  In both cases, the shape and position of the finite-frequency peak obtained in the fully quantum FTLM treatment coincide with the result of the static disorder problem at all frequencies $\omega \gtrsim \omega_0$, i.e. whenever the electrons are driven at a frequency that is faster than the bosons. 
The compelling agreement observed on the localization-induced peak demonstrates that thermally populated bosons are able to localize the electronic wavefunction even though there is no explicit random term in Eq. (\ref{eq:Holstein}). The fact that the conductivity is instead not completely suppressed when  $\omega < \omega_0$ indicates that the quantum interference processes at play here are only transient,
being eventually destroyed at  times longer than $\omega_0^{-1}$.  A detailed discussion of the d.c. limit $\omega\to 0$ is provided in the second part of this manuscript.

\paragraph{Conditions for transient localization and one-parameter scaling.---}
Localization originates from quantum interference effects not contained
in semi-classical descriptions of electron transport. Remarkably, the failure in capturing this phenomenon does not spare even  modern  sophisticated treatments  such as dynamical mean field theory (DMFT) or other large-N based approaches \cite{SYK-review} that
rely on self-averaging assumptions and therefore  disregard non-local interferences by construction; in diagrammatic terms, current vertex corrections are ignored \cite{Vucicevic-vtx}. This is illustrated in Figs. \ref{fig:spectra-vsT}(b) and (c), where alongside the FTLM and static boson results we report the optical spectra calculated within DMFT \cite{Fratini-PRL09}, showing no sign of a DDP.

\begin{figure}[h]
        \includegraphics[width=8.4cm]{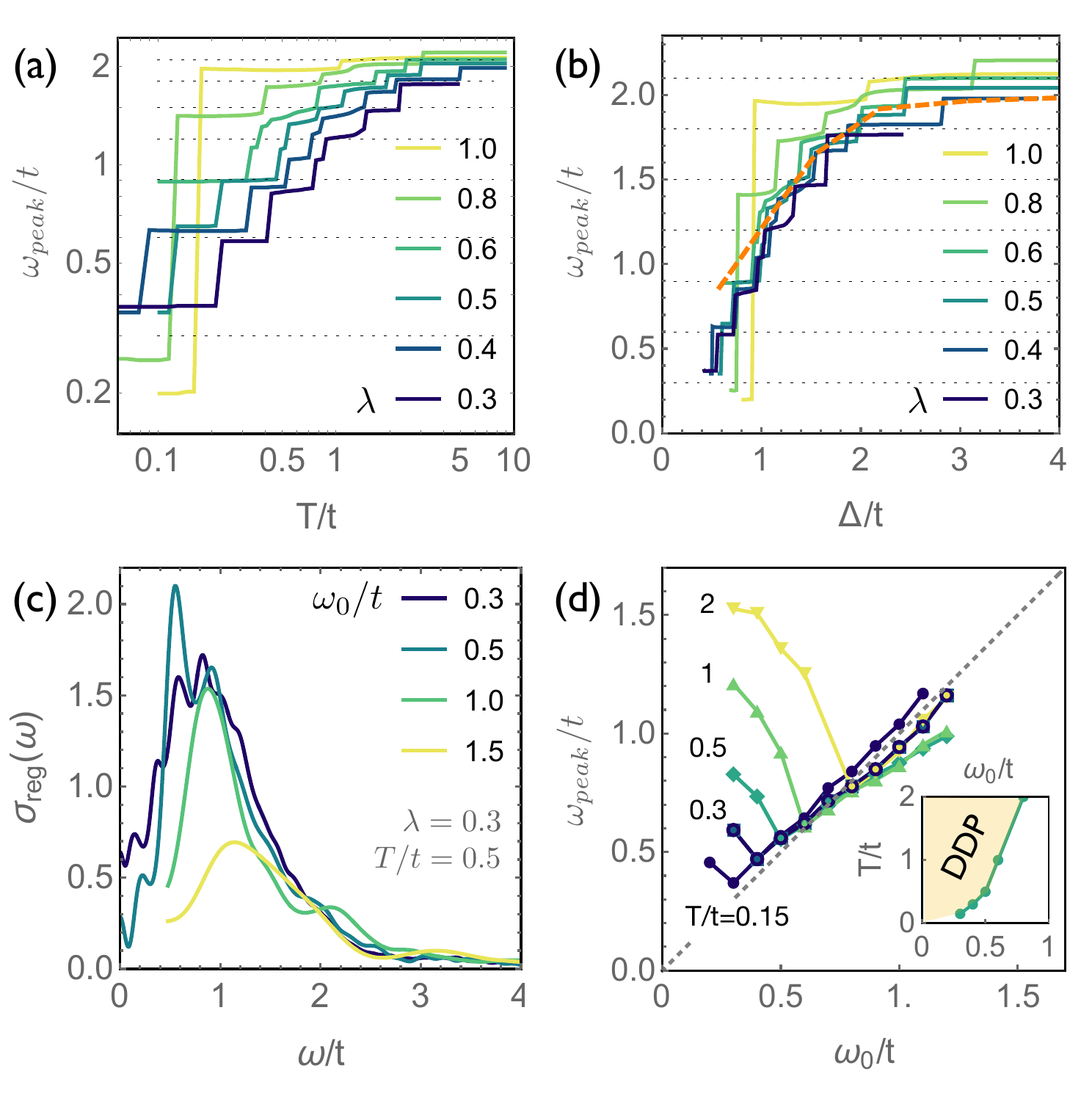}
\caption{ 
Peak position (a) as function of $T$ and (b) as a function of  self-generated disorder $\Delta$, for  $\omega_0=0.3$ and different values of $\lambda$. Dotted lines indicate multiples of $\omega_0$.  The dashed line is the result from the static boson limit for $\lambda=0.3$.
(c) Spectra at $\lambda=0.3$, $T/t=0.5$ and different $\omega_0$, with $\delta=0.15 \omega_0$. The low frequency part becomes inaccessible at large $\omega_0$, see SM.
(d) Peak position vs. $\omega_0$ at different $T$; the dashed line is $\omega_{peak}=\omega_0$. The inset shows the temperature range of existence of the DDP as extracted from the data in the main panel.}
\label{fig:peak-dependence}
\end{figure}

To analyze the phenomenon further, we report in Fig. \ref{fig:peak-dependence}(a) the evolution of the peak frequency with temperature at different values of the electron-boson coupling strength. In all cases the peak frequency increases monotonically with $T$, and it also shows an overall increase with $\lambda$ at fixed $T$. This behavior can be collapsed onto a single curve when expressed   as a function of the variance $\Delta$ of the emergent bosonic disorder, which is illustrated in Fig. \ref{fig:peak-dependence}(b). The superimposed steps observed at multiples of $\omega_0$ originate from the multi-boson fine structure seen in Fig.  \ref{fig:spectra-vsT}(a).

Implicit in the arguments given in the preceding paragraphs, two conditions must be met for the emergence of a localization-induced DDP. (i) The bosons must be largely --- and incoherently --- populated, which requires that $T$ is larger than the Debye temperature of the bosons. For weak interactions this translates into $k_BT\gtrsim \hbar\omega_0$, but the DDP can in principle be sustained down to lower temperatures if the boson frequency is itself renormalized by interactions, which is known to happen when interactions are strong  \cite{SciPost}.  (ii) The dynamics of boson-induced disorder, governed by the timescale $\omega_0^{-1}$, must be slower than the characteristic time of localization, $\omega_L^{-1}$, i.e. $\omega_L > \omega_0$: only in this case the localization processes can build up in the disordered environment created by the bosons, that the electrons will see as quasi-static.

To illustrate these conditions, Fig. \ref{fig:peak-dependence}(c) shows the evolution of the optical absorption upon varying $\omega_0$ at a fixed temperature $T=0.5t$. For $\omega_0<T$ the spectrum displays a disorder-induced DDP, here located  at $\omega_L\simeq 0.8t$ (same data as in Fig. \ref{fig:spectra-vsT}). Upon increasing $\omega_0$ (reducing $T/\omega_0$)  the DDP initially softens as boson coherence builds up, effectively reducing the amount of thermal disorder. As soon as $\omega_L$ hits $\omega_0$, the peak bounces back and hardens again, following $\omega_{peak}=\omega_0$: the disorder-induced DDP has disappeared in favor of  a more conventional single-boson peak.  The same evolution is actually observed at all temperatures, starting from different initial values of $\omega_L$ in the limit $\omega_0\to 0$ (Fig. \ref{fig:peak-dependence}(d)).

\paragraph{Suppression of the conductivity and validation of the TL formula.---}
\begin{figure}[h]
       \includegraphics[width=8.4cm]{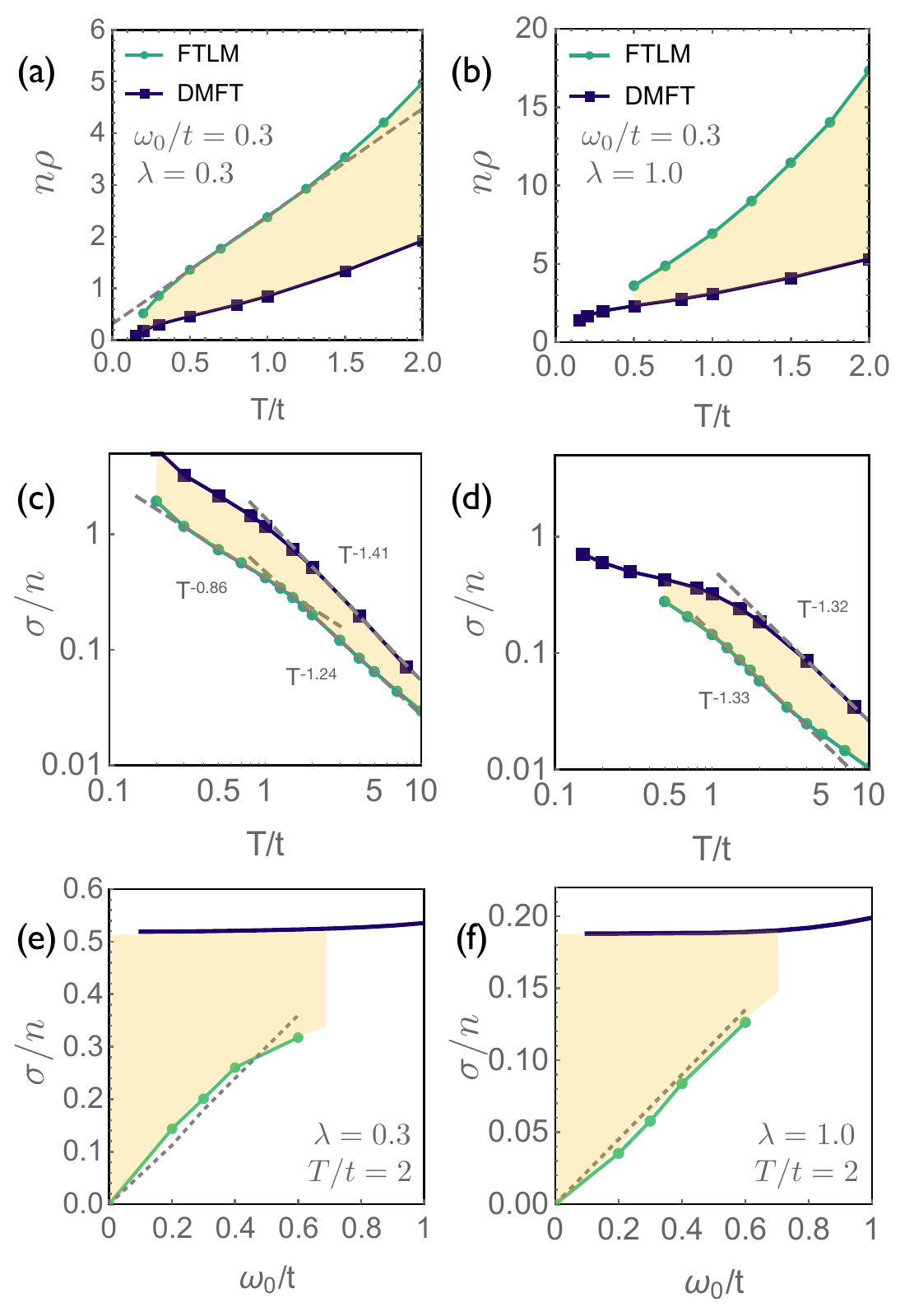}
\caption{
(a,b) Temperature dependence of the resistivity per particle, calculated by extrapolating the FTLM data to $N_{bos}=\infty$ at $N_s=4$ for $\omega_0/t=0.3$,  $\lambda=0.3$ (a) and  $\lambda=1.0$ (b) ($\delta=0.08t$).    The shaded area represents the increase in resistivity caused by quantum localization processes. (c,d) conductivity per particle, plotted in $\log$ scale to highlight the power-law dependence at high T (labels).
(e,f) Dependence of the conductivity on the bosonic scale $\omega_0$ at $T=2$ (e) for  $\lambda=0.3$  and (f) for $\lambda=1.0$. 
The dashed line is the prediction of the transient localization formula based on the observed DDP position (see text).}
    \label{fig:mobility-vsT}
\end{figure}
We finally come to the key point concerning charge transport: the suppression of low frequency spectral weight that accompanies the formation of the displaced Drude peak extends down to $\omega=0$, causing a suppression of the electrical conductivity (an increase in resistivity) as compared to semi-classical estimates. This is illustrated in Fig. \ref{fig:mobility-vsT}, showing the temperature dependence of the resistivity at $\omega_0/t=0.3$ for $\lambda=0.3$ (a) and  $\lambda=1.0$ (b). Each data point has been obtained by extrapolating to the $N_{bos}\to \infty$ limit for fixed cluster size, and then checking that the result is independent on $N_s$ (cf. SM), being therefore representative of the thermodynamic limit. The calculated resistivity is systematically larger than the DMFT result, due to quantum localization corrections.

Fig. \ref{fig:mobility-vsT}(a) reveals an approximately T-linear resistivity in the moderate electron-boson coupling regime (dashed line). The log scale plots in panels (c,d) provide more precise estimates of the power-law dependence of the conductivity. Comparison with DMFT shows that localization corrections amount to a mostly $T$-independent suppression factor, only weakly affecting the power-law exponent.

We now show  how the reduction of the conductivity depends on the bosonic scale $\omega_0$. Charge diffusion is  completely suppressed when $\omega_0=0$ (full Anderson localization) while in the opposite limit, localization effects are destroyed when $\omega_0 > (T,\omega_L)$ as discussed already.  The conductivity in the available range of $\omega_0$, reported in Fig.  \ref{fig:mobility-vsT}(e,f), is  compatible with interpolating between full localization and the semi-classical DMFT result. The suppression factor introduced by localization is not only $T$-independent, as shown in the preceding paragraph, but also $\lambda$-independent, being therefore primarily governed by the ratio $\omega_0/t$.

Finally, the numerical data presented here quantitatively validate the relaxation time approximation that was originally introduced to describe charge transport  in the transient localization regime \cite{PRBR11}. Following early ideas by Thouless \cite{Thouless}, it was argued that in this regime the mobility $\mu=\sigma/ne$ should follow $\mu_{TL}=(e/k_BT)L^2/2\tau_{0}$, with $\tau_{0}$ the typical timescale of the bosonic fluctuations,  instead of the usual Drude expression $\mu=e\tau/m$. By extracting the localization length $L=\sqrt{2t/\omega_L}$ from the position of the DDP we find that the TL formula is in excellent agreement with the numerical results if we set $\tau_{0}^{-1}=\alpha \omega_0$ with $\alpha=1/2.2$ as in Ref. \cite{Fratini20}.

\paragraph{Concluding remarks.---}
We would like to conclude by highlighting two concepts that could be of relevance in quantum materials and that follow directly from the results presented in this work. First, our findings unequivocally demonstrate that localization effects can arise as a general consequence of self-generated disorder. This was illustrated here in a paradigmatic model for electron-boson interactions, yet the potential implications of the phenomenon are  broader: collective excitations resulting from many-body interactions and critical modes near quantum phase transitions are two examples of bosonic degrees of freedom that are commonly found in quantum matter, that are intrinsically soft, and could therefore cause quantum localization  analogous  to that described here. We advocate that any emergent randomness in interacting electron systems \cite{SYK-review} should be taken at face value, especially in low-dimensions, without disregarding the quantum processes generally associated with Anderson localization. The latter would instead be entirely missed by self-averaging theories such as dynamical mean field theory  \cite{optcondDMFT,Fratini-PRL09,Vucicevic-vtx} and other large-N  treatments \cite{SYK-review}.

Second, it is often assumed that quantum localization corrections are only relevant at very low temperatures, while they can be ignored otherwise.
Our results demonstrate that the opposite is true whenever the randomness causing localization is self-generated: in this case the equipartition principle dictates that random fluctuations will grow with temperature, overcoming the dephasing effects driven by the temperature itself. 
Theory shows that  localization processes can develop from electron-boson interactions  rather than being washed out by them. From an experimental standpoint,  the widespread observation of displaced Drude peaks in clean quantum materials \cite{SciPost23,Takenaka,Pustogow-NatComm,Lupi,Uykur,Lee02,Wang04,SciPost23} could be hinting towards a  broader relevance of the self-generated localization mechanism established here.

\acknowledgments
S.C. acknowledges funding from  NextGenerationEU National Innovation Ecosystem grant ECS00000041 - VITALITY - CUP E13C22001060006.

\bibliography{TLFTLM}

\clearpage

\appendix
\begin{center}
    \Large{Supplementary information}    
\end{center}

\pagenumbering{roman}

\setcounter{figure}{0}
\renewcommand{\figurename}{Fig.}
\renewcommand{\thefigure}{S\arabic{figure}}

\setcounter{equation}{0}
\renewcommand{\theequation}{S.\arabic{equation}}

\section{Method}
We solve the model Eq. (\ref{eq:Holstein}) using the finite temperature Lanczos method (FTLM) \cite{FTLM_1,prelovsek} for a single electron on finite-size chains of length $N_s$. This method provides a formally exact high-temperature expansion for both the thermodynamic quantities and dynamic correlation functions down to temperatures of the order of a fracion of the unit energy scale of the problem, $t$ in Eq. (1).  For the specific problem at hand, the necessary truncation of the infinite  Hilbert space of the bosons to a maximum total number of quanta $N_{bos}$ also sets a constraint on the highest temperatures attainable (see the Section {\em Hamiltonian construction} below).

In practice, the FTLM  evaluates thermodynamic quantities  by employing the concept of trace estimators in a reduced Krylov space of size $M$, much smaller than the original size of the Hilbert space, $N_{Hilb}$ \cite{FTLM_trace}. 
The thermodynamic averages are then replaced by  averages over random initial vectors $|r\rangle$, which can be viewed as superpositions of an orthogonal set of basis vectors. While one single random vector already provides reliable results at high temperatures, a larger number of random extractions is required as the temperature is lowered \cite{prelovsek}. In this work we use averages over $40$ random vectors.


We study clusters of size up to $N_s=6$. Finite size effects are minimized by averaging over $N_\phi$ twisted boundary conditions (TBC) \cite{TBC_1,Driscoll}, increasing the effective number of allowed momenta from the nominal value $N_s$ to  $N_s \times N_\phi$. Here we use a regular grid of fluxes connecting the nominal momenta of the original cluster. TBC sampling permits us to obtain reliable results down to the weak coupling regime, where the discrete nature of the non-interacting spectrum on small clusters is most critical. Throughout this work we use $N_\phi=40$, performing twist averages and random vector averages simultaneously. 

Regarding the size of the Krylov space  we use $M=300$, allowing us to obtain  excitation spectra that are sufficiently dense for our purposes  (see the Section {\em Optical conductivity} below).

\subsection{Basis representaion} 
As any other exact diagonalization method, using FTLM requires the construction of the basis for the system. Basis construction plays a crtical role in time scaling and memory storage. The basis of the Holstien model consists of two parts: the electron part and the boson part. The electron part can be encoded using the site representation, which describes electronic orbitals at site i. The boson part can be represented by the boson number at each site. The combined information can be encoded in a one-dimensional array of length $2N_s$, where the first $N_s$ elements represent the location of the electron, and the second $N_s$ elements represent the number of bosons at each site.
A general example of such vector will be:
 \begin{equation}
    \centering
    \ket{\psi}=\ket{001000}\bigotimes\ket{145280}
 \end{equation}
The left ket represents the state of a 6-site chain where the electron is localizad on the third site,  which can be easily viewed as a binary number representation. The right ket corresponds to the boson numbers at each site.

Constructing the basis from the  representation discussed above requires counting all the possible electron locations over the chain of length $N_s$ and all the possible configurations of  $N_{bos}$  bosons over $N_{sites}$. This is easily evaluated through the stars and bars method and amounts to $\frac{(N_s+N_{bos}-1)!}{N_s!(N_{bos}-1)!}$. The total size of the Hilbert space is therefore $N_{Hilb}(N_s,N_{bos})=N_s \frac {(N_s + N_{bos}-1  )!}{(N_{bos}-1  )!  N_s!} =  \frac{(N_s + N_{bos}-1  )!}{(N_{bos}-1  )!  (N_s -1)!} $.


\subsection{Memory and size limitations}
The available memory limits the attainable cluster size to Mem $\sim 32 M \times N_{Hilb}(N_s,N_{bos})$. For $M=300$ and $256$ Gb of memory,  this value cannot exceed $\sim 10^7$ states, which corresponds to a maximum of $N_{bos}\simeq 30$ for $N_s=6$ (number of bosons per site $N_{bos}/N_s=5$), $N_{bos}\simeq 45$ for $N_s=5$ ($N_{bos}/N_s=9$) and $N_{bos}\simeq 90$ for $N_s=4$ ($N_{bos}/N_s>20$). 

The number of bosons per site controls the maximum temperature that can be attained, $T_{max} \sim (N_{bos}/N_s) \omega_0$. Sufficiently large values of $N_{bos}$ are required to be able to thermally populate an appropriate number of boson quanta $\gg T/\omega_0$ on each site, in order to recover the high-temperature limit of the Bose-Einstein distribution. For example, obtaining reliable results at $T/t=2$ for $\omega_0/t=0.3$ requires $N_{bos}/N_s\gg 7 $, which can only be attained for the $N_s=4$ size cluster.

\section{Optical conductivity}

\begin{figure}[h]
       \includegraphics[width=8.4cm]{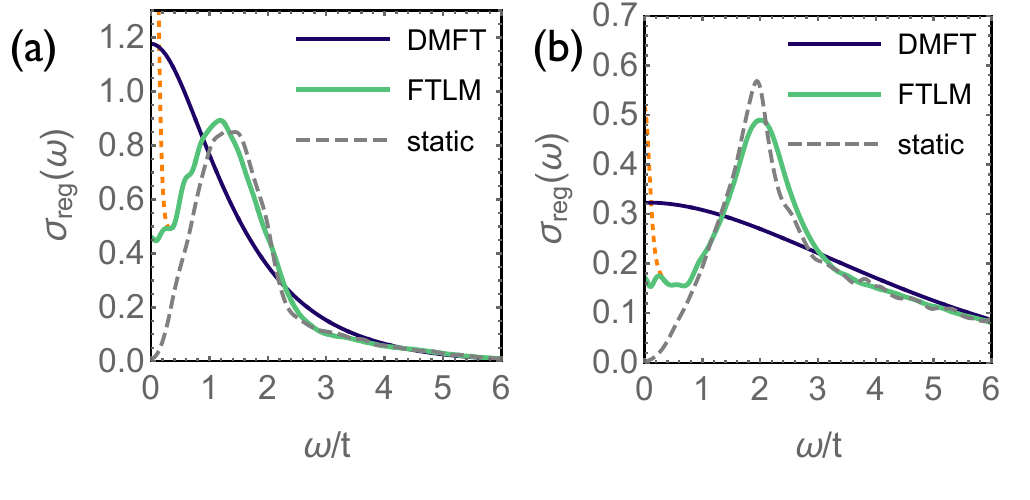}
\caption{Same as Fig. 1(b) and (c) of the main text, illustrating the spurious finite-size contribution at $\omega\approx 0$ for $\omega_0/t=0.3$,  $\lambda=0.3$ (a) and $\lambda=0.3$ (b).}
    \label{fig:regular}
\end{figure}

The optical conductivity is calculated within FTLM following the derivation provided in Ref. \cite{prelovsek}.
We write the optical conductivity as $\sigma(\omega)=\sigma_{reg}(\omega)+\sigma_0(\omega)$, separating explicitly the regular part at finite frequencies from the singular part arising from the charge stiffness at $\omega=0$, $\sigma_0=D \delta(\omega)$. The latter must vanish at all temperatures in the thermodynamic limit in the absence of persistent currents. However, the combination of the finite system sizes used, allowing for nonzero $D$, and the statistical uncertainties inherent to the FTLM method, redistributing the delta-function over several inelastic transitions, is responsible for the emergence of a narrow contribution at $\omega\approx 0$, as illustrated in Fig. \ref{fig:regular} (dashed line). In order to obtain results representative of the thermodynamic limit, we exclude this spurious contribution through fitting.

The regular part itself is constituted of a discrete set of delta-functions, due to finite system size.
Using $M$ states in the Krylov space for Lanczos diagonalization gives rise to $O(M)$ levels in the optical absorption spectrum at low temperature, where only transitions from the ground state are allowed, to a much more favorable $O(M^2)$ at high temperatures. Denoting $W$ the typical width $W$ of the electronic excitation spectrum this implies a typical level spacing of at best $2\times W/M\simeq 2\times 4 t / 300 \sim 0.03 t$ at high $T$ $2\times W/M^2\simeq 2\times 4 t / 300^2 \sim 10^{-4} t$ at high $T$ (the factor of $2$ accounts for positive and negative energies, $4t$ is the non-interacting bandwidth). In order to obtain smooth spectra from the Lanczos diagonalization data we therefore apply a gaussian filter of width $\delta$. Fig.1(a) and Fig.2(c,d) in the manuscript are obtained using  $\delta=0.15 \omega_0$. This  accounts for the increase in $W$ with $\omega_0$ while at the same time preserving the fine structure induced by $\omega_0$ overtones. For the typical $\omega_0=0.3t$ used in most of the results presented, this yields a filter $\delta=0.045t$. Fig.1(b,c),  Fig.2(a,b) and Fig.3 are instead obtained with $\delta=0.08t$.

For the removal of the spurious contribution at $\omega\approx 0$ we first apply the Gaussian broadening $\delta$ and then fit the absorption  up to $\omega=0.5t$ as $\sigma(\omega)=\sigma_{reg}(\omega)+g(\omega)$, taking the lowest order symmetric contribution $\sigma_{reg}(\omega)=\sigma_{reg}(0)+A\omega^2$ and $g(\omega)$ itself a Gaussian. We then remove the fitted $g(\omega)$ to obtain the regular spectrum at all frequencies, as shown in Fig. \ref{fig:regular}. The result is only marginally affected by the initial choice of the broadening parameter $\delta$. The procedure described here cannot be used at large $\omega_0/t$, outside of the transient localization regime, where a narrow Drude peak is expected to arise \cite{optcond08} and would be indistinguishable from the supurious stiffness. For this reason for all cases $\omega_0/t \ge 0.6$ we only show in the manuscript the finite frequency part of the optical spectra and do not attempt the extrapolation to the d.c. limit.

\section{Finite size scaling}

\begin{figure}[th]
       \includegraphics[width=8.4cm]{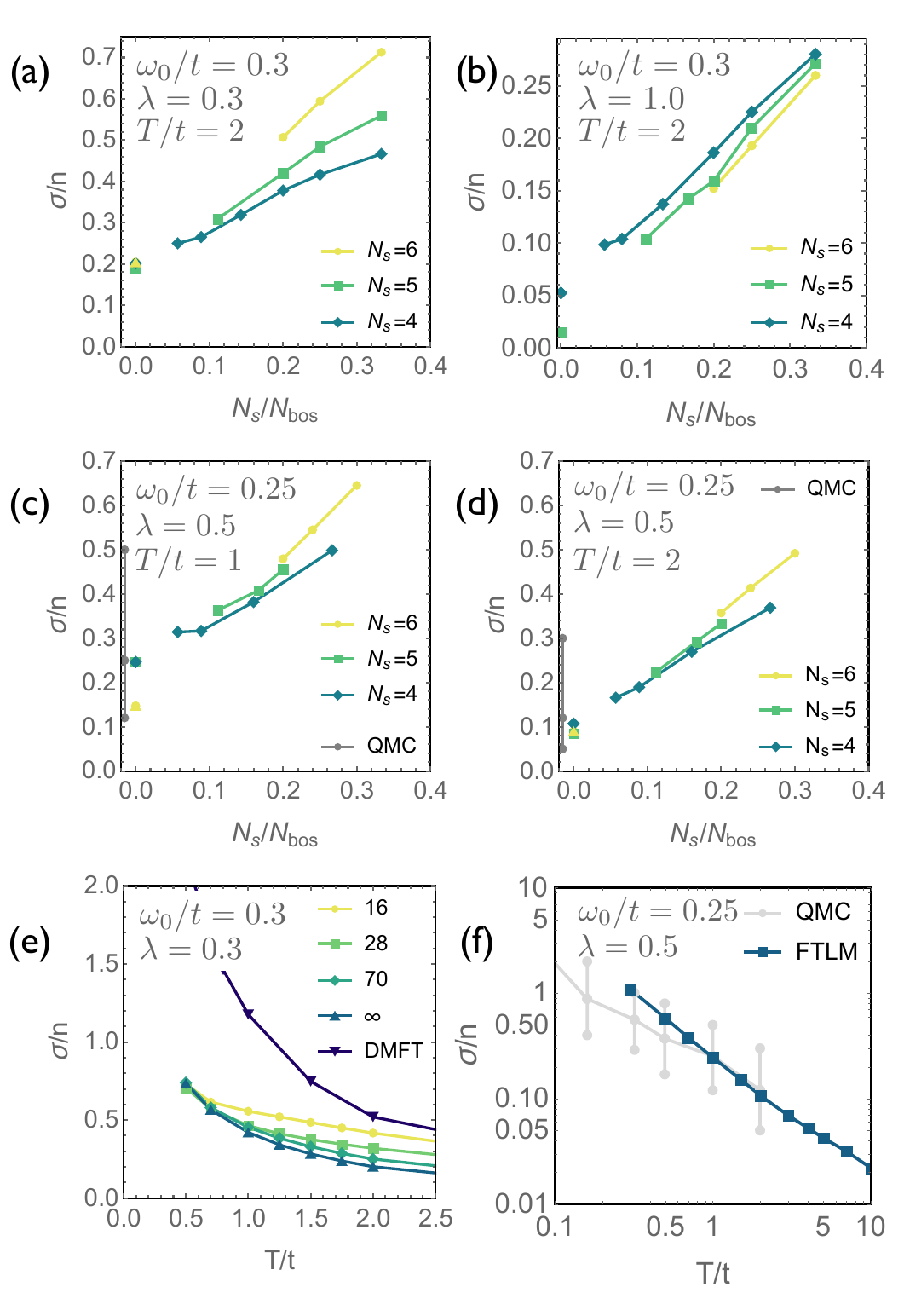}
\caption{Illustrations of finite-size scaling and convergence as described in the text.}
    \label{fig:sizescaling}
\end{figure}

After removal of the singular part, the d.c. conductivity per particle (electron mobility) can be obtained as the $\omega\to 0$ limit of  $\sigma_{reg}(\omega)$. Fig. \ref{fig:sizescaling}(a) shows a representative example of the  convergence of the conductivity to the thermodynamic limit, for $\omega_0/t=0.3$, a moderate $\lambda=0.3$ and $T/t=2$.  The graph shows how the extrapolated conductivity evolves as a function of the number of bosons per site $N_{bos}/N_s$ to the limit $N_{bos}\to \infty$ ($N_s/N_{bos}\to 0$). Different curves obtained for different system sizes $N_s$  all extrapolate to the same value, which is therefore representative of the thermodynamic limit $N_s\to \infty$, $N_{bos}\to \infty$. The relative uncertainty increases at for strong interactions, $\lambda=1$ (panel (b)), where the conductivity itself is suppressed by polaronic effects, reducing the signal to noise ratio.  

Panels (c,d) show results at intermediate $\lambda=0.5$, benchmarked against the quantum Monte Carlo (QMC) results of Ref. \cite{Mishchenko} (gray errorbars on the left axis). The agreement of our extrapolated data is excellent for both $N_s=4$ and $N_s=5$. The limited maximum number of bosons at $N_s=6$ makes the extrapolation to $N_{bos}\to \infty$ less precise, yet providing results that fall into the errorbar of QMC. The full temperature-dependence for this parameter set and the comparison with QMC are shown in panel (f).

Finally, Fig. \ref{fig:sizescaling}(e) shows the temperature dependence of the conductivity obtained at different values of $N_{bos}$ (legends) for $N_s=4$ as well as the curve extrapolated to $N_{bos}\to \infty$, for the same parameters as in panel (a). The DMFT result is always larger than the FTLM result, showing that the effect of localization (vertex) corrections discussed in the manuscript is not a finite-size artifact.

\end{document}